\documentclass[epj-spec]{svjour}
\usepackage{graphicx,color}
\usepackage{amsmath}
\usepackage{amssymb}
\usepackage{bm}

\newcommand{\Tr}{\mathop{\text{Tr}}\nolimits}

\begin{document}
\title{XX model on the circle}
\author{A. De Pasquale\inst{1}
\and G. Costantini\inst{1,2}
\and P. Facchi\inst{3,2}
\and G. Florio \inst{1,2}
\and S. Pascazio\inst{1,2}
\and K. Yuasa\inst{4,5}
}
\institute{Dipartimento di Fisica, Universit\`a di Bari, I-70126 Bari, Italy \and Istituto Nazionale di Fisica Nucleare, Sezione di Bari, I-70126 Bari, Italy \and Dipartimento di Matematica, Universit\`a di Bari, I-70125 Bari, Italy \and Waseda Institute for Advanced Study, Waseda University, Tokyo 169-8050, Japan \and Department of Physics, Waseda University, Tokyo 169-8555, Japan}
\abstract{We diagonalize the XX model with a finite number
of spins and periodic boundary conditions. We solve for the ground
state, focus on the rapidity of the convergence to the thermodynamic
limit and study the features of multipartite entanglement.
}
\maketitle

\section{Introduction}
Recent advances in quantum information processing motivated a
widespread interest for spin systems as models for quantum computers
\cite{nielsen}.
Spin chains are often studied, in particular in the thermodynamic
limit \cite{sachdev,lieb,pfeuty}, where one can neglect the
contributions deriving from the finite size of the system, that
scale like $O(1/N)$,
with $N$ the number of spins. However, due
to experimental and theoretical difficulties, such as decoherence
and/or imperfections in the quantum hardware, nowadays it is only
possible to assemble and control the interaction of a small number
of qubits (always less than $10$). For this reason, it is extremely
important to take into account all finite-size effects in the system
Hamiltonian.

The investigation of the last few years has focused on entanglement
\cite{sarorev} in diverse finite-size models, by means of direct
diagonalization \cite{tagliafinita}. These studies were boosted by
the recent discovery that entanglement can detect the presence of
quantum phase transitions \cite{QPT}. In this article we will study
the finite-size 1D quantum XX model with periodic boundary
conditions. We will exactly diagonalize the Hamiltonian by a
Jordan-Wigner transformation followed by a deformed Fourier
transform. Some of the results we will obtain were alluded to in the
seminal article by Lieb \textit{et~al.}\ \cite{lieb} and were
discussed by \v{S}telmachovi\v{c} and Bu\v{z}ek \cite{SB} and very
recently by Canosa and Rossignoli \cite{canosarossignoli}. We will
focus here on the features of the ground state and evaluate the
``forerunners" of the critical points, as the quantum phase
transition (QPT) occurs only in the thermodynamical limit. By
introducing a finite-size parameter, we will show that the
thermodynamical limit is very rapidly approached, so that a $10$
qubit system already represents a very good approximation. Finally,
we shall analyze the multipartite entanglement features of the
finite-size ground state.

\section{Exact diagonalization of the finite size XX model}
\subsection{Exact diagonalization of the Hamiltonian}
\begin{figure}
\begin{center}
\includegraphics[width=0.3\columnwidth]{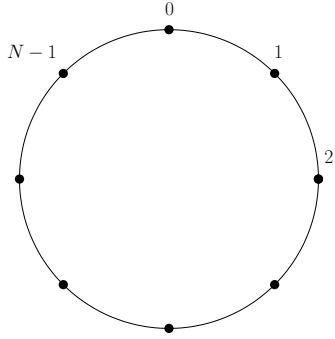}
\end{center}
\caption{A spin chain with periodic boundary conditions.}
\label{fig:Spin chain}
\end{figure}

The XX model for a collection of $N$ qubits on a chain is described by the Hamiltonian
\begin{equation}\label{eq:hamitonianoXXPauli}
H=-J\sum_{i=0}^{N-1}\left(
g\sigma_i^z + \frac{1}{2}\sigma_i^x
\sigma_{i+1}^x + \frac{1}{2}\sigma_i^y \sigma_{i+1}^y
\right),
\end{equation}
where $J$ is a constant with dimensions of energy and $g$ a
dimensionless parameter proportional to the transverse magnetic
field. We consider periodic boundary conditions (cyclic chain, see
Fig.\ \ref{fig:Spin chain})
\begin{equation}
\bm{\sigma}_0 = \bm{\sigma}_N.
\end{equation}

The usual diagonalization scheme is a two-step procedure based on
the Jordan-Wigner (JW) transformation followed by a Fourier
transform. The use of the JW transformation is based on the
observation that there exists a unitary mapping between the Hilbert
space of $N$ qubits $(\mathbb{C}^2)^{\otimes N}$ and the
antisymmetric Fock space $\mathcal{F}_-(\mathbb{C}^N)$ of spinless
fermions on $N$ orbitals.  We can define annihilation and creation
operators for $\mathcal{F}_-$,
\begin{equation}
c_i=\Biggl(\prod_{0\leq j<i}\sigma^z_j\Biggr)\,\sigma_i^-
=e^{\text{i}\pi\mathbf{n}_{i\downarrow}}\sigma_i^-,\quad
c_i^\dag=\Biggl(\prod_{0\leq j<i}\sigma^z_j\Biggr)\,\sigma_i^+
=e^{\text{i}\pi\mathbf{n}_{i\downarrow}}\sigma_i^+,\quad
\forall i\in\mathbb{Z}_N,
\label{eq:defcc}
\end{equation}
where $\sigma_i^\pm=(\sigma_i^x\pm\text{i}\sigma_i^y)/2$,
$\mathbb{Z}_N=\{0,1,\ldots,N-1\}$,
and $\mathbf{n}_{i\downarrow}$ is the number operator counting the holes
between $0$ and $i-1$,
\begin{equation}
\mathbf{n}_{i\downarrow}
=\sum_{j=0}^{i-1}(1-\sigma_j^+\sigma_j^-)
=\sum_{j=0}^{i-1}\sigma_j^-\sigma_j^+.
\end{equation}
It can be shown that the JW operators satisfy anticommutation
relations
\begin{equation}
\{c_i,c_j\}=0,\quad
\{c_i^\dag,c_j^\dag\}=0,\quad
\{c_i,c_j^\dag\}=\delta_{ij},\quad
\forall i, j \in\{0,1,\ldots,N-1\},
\end{equation}
whereas the corresponding Pauli operators anticommute only on
site:
\begin{equation}
\{\sigma_i^\pm,\sigma_j^\pm\}=0\quad\text{for}\quad i=j,
\quad
[\sigma_i^\pm,\sigma_j^\pm]=0\quad\text{for}\quad i \neq j.
\end{equation}

The periodic boundary conditions assigned to the cyclic chain do not
hold in general for the fermionic operators $c_i$; indeed, by
prolonging (\ref{eq:defcc}) up to $i=N$ we would obtain
\begin{equation}
c_0=\sigma_0^-,\quad
c_N=e^{\text{i}\pi\mathbf{n}_\downarrow}\sigma_N^-
=e^{\text{i}\pi\mathbf{n}_\downarrow}
\sigma_0^-=e^{\text{i}\pi\mathbf{n}_\downarrow}c_0,
\end{equation}
where $\mathbf{n}_\downarrow=\mathbf{n}_{N\downarrow}$ is the number operator that counts the
total number of spins down (and the number of holes), from $0$ to $N-1$. As a consequence, the Hamiltonian
written in terms of the JW operators is characterized by the
presence of a boundary term:
\begin{equation}\label{eq:hamiltonianoXXjordan-wigner}
H=-J\,\Bigg[
\sum_{j=0}^{N-1}g(1-2c_jc_j^\dag)+
\sum_{j=0}^{N-2}(c_jc_{j+1}^\dag+c_{j+1}c_j^\dag)
+(e^{\text{i}\pi(\mathbf{n}_\downarrow+1)}-1)
(c_{N-1}c_0^\dag+c_0c_{N-1}^\dag)\Bigg].
\end{equation}
In the thermodynamic limit \cite{lieb,pfeuty} the XX Hamiltonian is
diagonalized by introducing the discrete Fourier transform
\begin{equation}
c_j=\frac{1}{\sqrt N}\sum_{k=0}^{N-1}e^{\frac{2\pi
\text{i}kj}{N}}\hat{c}_k, \quad \forall j\in\mathbb{Z}_N
\label{eq:aaa}
\end{equation}
and neglecting the last term in
(\ref{eq:hamiltonianoXXjordan-wigner}), since its contribution
scales like $O(1/N)$. We intend to evaluate the corrections due to
the finite size of the chain.

The main difficulty introduced by the boundary term in the
Hamiltonian (\ref{eq:hamiltonianoXXjordan-wigner}) is that it breaks
the periodicity of the JW operators, due to the arbitrary dependence
of the phase $e^{\text{i}\pi \mathbf{n}_{i\downarrow}}$ on the
ordering of the spin on the circle. This phase clearly depends on
the state the Hamiltonian $H$ is applied to. However, notice that
the parity of the spin-down number operator is conserved (although
not so the number operator itself)
\begin{equation}\label{eq:commparity-hamiltonian}
[e^{\text{i}\pi \mathbf{n}_\downarrow},H]=0.
\end{equation}
Therefore, let us consider the parity operator
\begin{equation}
\mathcal{P}= e^{\text{i}\pi (\mathbf{n}_\downarrow+1)}
\end{equation}
and the spectral decomposition of the number operator
$\mathbf{n}_\downarrow$,
\begin{equation}
\mathbf{n}_\downarrow=\sum_{n_\downarrow=0}^{N}n_\downarrow|n_\downarrow\rangle\langle
n_\downarrow|,
\end{equation}
where $|n_\downarrow\rangle$ is the eigenstate of
$\mathbf{n}_\downarrow$ belonging to the eigenvalue $n_\downarrow$. In this
basis the parity operator can be written as:
\begin{equation}
\mathcal{P}=e^{\text{i}\pi
(\mathbf{n}_\downarrow+1)}\sum_{n_\downarrow=0}^N|n_\downarrow\rangle\langle
n_\downarrow|=P_+ -P_-,
\end{equation}
where
\begin{equation}
P_+=\sum_{n_\downarrow\,\text{odd}}|n_\downarrow\rangle\langle
n_\downarrow|,\qquad P_-=\sum_{n_\downarrow\,\text{even}}|n_\downarrow\rangle\langle n_\downarrow|
\end{equation}
are the projection operators associated to the eigenvalues $\pm1$ of
$\mathcal{P}$, respectively. Due to Eq.\
(\ref{eq:commparity-hamiltonian}), the Hamiltonian preserves the
parity sectors and  can be decomposed as
\begin{equation}
H=P_+HP_+ + P_-HP_-=H^{(+)} + H^{(-)}.
\label{eq:decomposition}
\end{equation}
The analysis can then be separately performed in the two parity
sectors, where $\mathcal P$ acts as a c-number.

Let us define a deformed Fourier transform
\begin{equation}\label{inverse new fourier transform}
c_j=\frac{1}{\sqrt N}e^{\frac{2\pi
\text{i}\alpha_j}{N}}\sum_{k=0}^{N-1}e^{\frac{2\pi
\text{i}kj}{N}}\hat{c}_k ,
\end{equation}
where we added a position dependent phase $\alpha_j$
($j\in\mathbb{Z}_N$ denoting the site). The anticommutation
relations for $\hat{c}_k$ still hold:
\begin{equation}
\{\hat{c}_k,\hat{c}_{k'}\}=0,\quad
\{\hat{c}_k^\dag,\hat{c}_{k'}^\dag\}=0,\quad
\{\hat{c}_k,\hat{c}_{k'}^\dag\}=\delta_{kk'}.
\end{equation}
The phase $e^{\frac{2\pi \text{i}\alpha_j}{N}}$ can be determined by
imposing that the last term of Eq.\
(\ref{eq:hamiltonianoXXjordan-wigner}), after Fourier transform, has
the same form of the other $N-1$ terms: namely we require that the
following two expressions,
\begin{align}
c_jc_{j+1}^\dag
&=e^{\frac{2\pi
\text{i}\alpha_j}{N}}e^{-\frac{2\pi\text{i}\alpha_{j+1}}{N}}\frac{1}{N}\sum_{k=0}^{N-1}\sum_{k'=0}^{N-1}e^{\frac{2\pi\text{i}jk}{N}}
e^{-\frac{2\pi\text{i}(j+1)k'}{N}}\hat{c}_k\hat{c}_{k'}^\dag
\label{c_jc_(j+1)^+}
\intertext{and}
e^{\text{i}\pi({\mathbf{n}_\downarrow}+1)}c_{N-1}c_0^\dag
&=e^{\text{i}\pi({\mathbf{n}_\downarrow}+1)}e^{\frac{2\pi\text{i}\alpha_{N-1}}{N}}
e^{-\frac{2\pi\text{i}\alpha_0}{N}}\frac{1}{N}\sum_{k=0}^{N-1}\sum_{k'=0}^{N-1}e^{\frac{2\pi\text{i}(N-1)k}{N}}
\hat{c}_k\hat{c}_{k'}^\dag\nonumber\\
&=e^{\text{i}\pi({\mathbf{n}_\downarrow}+1)}e^{\frac{2\pi\text{i}\alpha_{N-1}}{N}}
e^{-\frac{2\pi\text{i}\alpha_0}{N}}\frac{1}{N}\sum_{k=0}^{N-1}\sum_{k'=0}^{N-1}e^{\frac{2\pi\text{i}(N-1)k}{N}}e^{-\frac{2\pi\text{i}Nk'}{N}}
\hat{c}_k\hat{c}_{k'}^\dagger
\label{c_(N-1)c_0^+},
\end{align}
have the same phase. It follows that
\begin{equation}\label{eq:phasecondition}
e^{\frac{2\pi\text{i}(\alpha_j-\alpha_{j+1})}{N}}=e^{\text{i}\pi({\mathbf{n}_\downarrow}
+ 1)}e^{\frac{2\pi\text{i}(\alpha_{N-1}-\alpha_0)}{N}}.
\end{equation}
The solution is given by
\begin{equation}
\alpha_j=j\alpha + \alpha_0,
\end{equation}
where $\alpha$ satisfies the equation
\begin{equation}\label{fase alpha}
e^{2\pi\text{i}\alpha}=e^{\text{i}\pi({\mathbf{n}_\downarrow}+1)}.
\end{equation}
On the other hand, the phase of the first site $\alpha_0$ is a free parameter. The two
solutions in the two parity sectors are
\begin{equation}
\alpha=
\begin{cases}
0\ \text{mod}\ N&\text{if}\quad \mathcal{P}=+1 \qquad
({n_\downarrow}\;\text{odd}),\\
\frac{1}{2}\ \text{mod}\ N &\text{if}\quad  \mathcal{P}=-1 \qquad ({n_\downarrow}
\;\text{even}).
\end{cases}
\label{alphadef}
\end{equation}
Summarizing, by using  the (sector dependent) deformed Fourier transform
\begin{equation}
c_j=\frac{1}{\sqrt N}e^{\frac{2\pi
\text{i}\alpha_0}{N}}e^{\frac{2\pi \text{i}\alpha
j}{N}}\sum_{k=0}^{N-1}e^{\frac{2\pi \text{i}kj}{N}}\hat{c}_k
\end{equation}
and substituting into Eq.\ (\ref{eq:hamiltonianoXXjordan-wigner}) we
obtain
\begin{equation}\label{eq:newhamdiag}
H=-2J\sum_{k=0}^{N-1}\left(\hat{c}_k^\dag
\hat{c}_k-\frac{1}{2}\right)\left[
g-\cos\!\left(
2\pi\frac{\alpha+k}{N}
\right)\right],
\end{equation}
where $\alpha$ is given by (\ref{alphadef}).
In this way, $H$ is diagonalized and the boundary term removed.
Therefore, the Hamiltonian is given by (\ref{eq:decomposition}) with
\begin{align}
H^{(+)}&=-2JP_+\sum_{k=0}^{N-1}\left(\hat{c}_k^\dag
\hat{c}_k-\frac{1}{2}\right)
\left[
g-\cos\!\left(2\pi\frac{k}{N}\right)\right]P_+,\\
H^{(-)}&=-2JP_-\sum_{k=0}^{N-1}\left(\hat{c}_k^\dag
\hat{c}_k-\frac{1}{2}\right)
\left[g-\cos\!\left(2\pi\frac{k}{N}+\frac{\pi}{N}\right)\right]P_-.
\end{align}

\subsection{Energy spectrum}
\begin{figure}[b]
\begin{center}
\includegraphics[width=0.4\columnwidth]{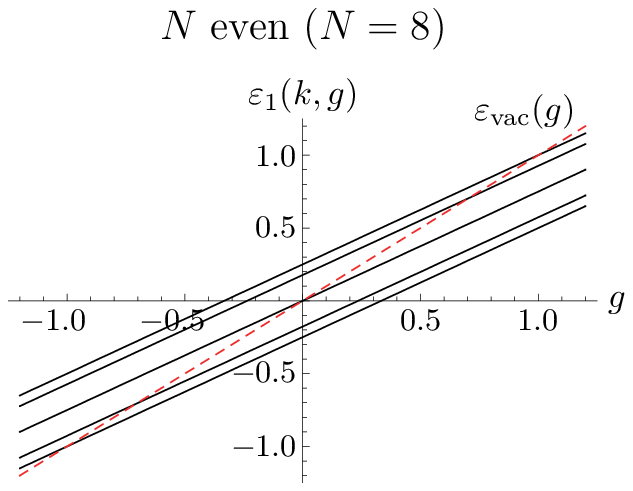}
\qquad
\includegraphics[width=0.4\columnwidth]{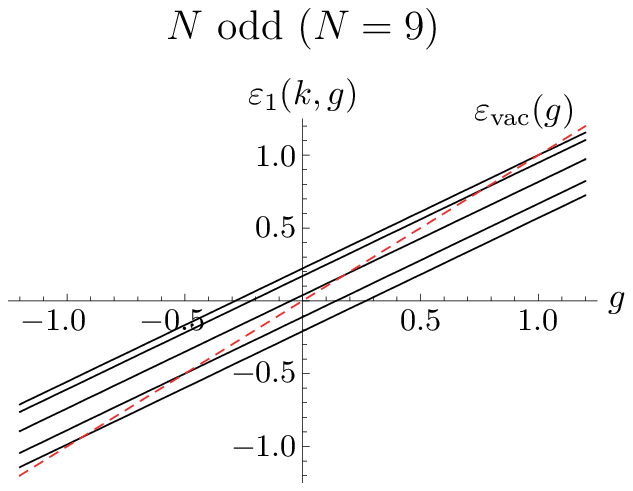}
\end{center}
\caption{Single particle spectra $\varepsilon_1(k,g)$ (solid lines) and vacuum density energies $\varepsilon_\text{vac}(g)$
(dashed lines). Different lines correspond to different
$k\in\mathbb{Z}_N$ according to (\ref{eq:singleenergy}).}
\label{fig:spettri singola particella}
\end{figure}

\begin{figure}
\begin{center}
\begin{tabular}{cc}
\includegraphics[width=0.4\columnwidth]{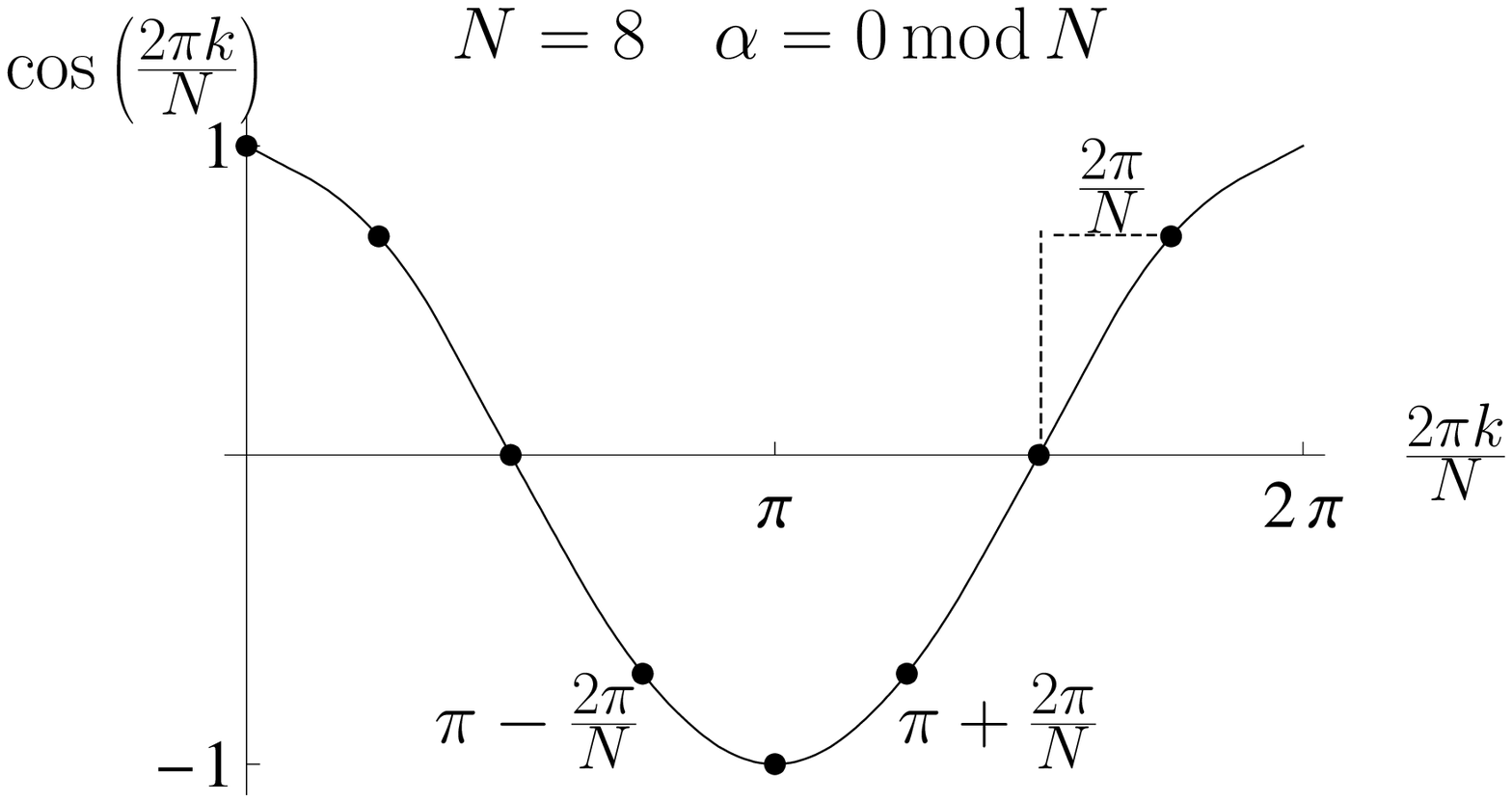}
&
\includegraphics[width=0.4\columnwidth]{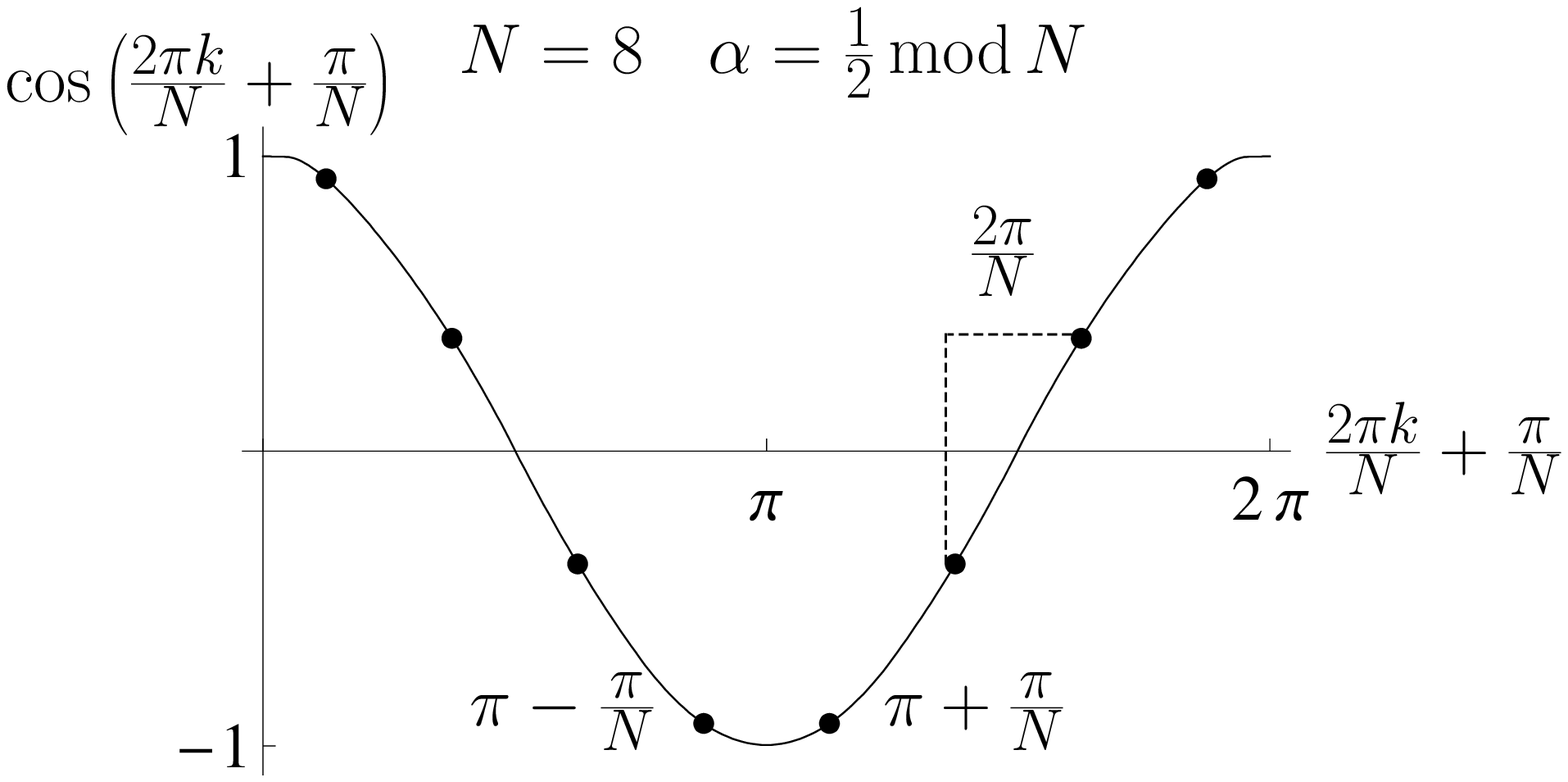}\\[5mm]
\includegraphics[width=0.4\columnwidth]{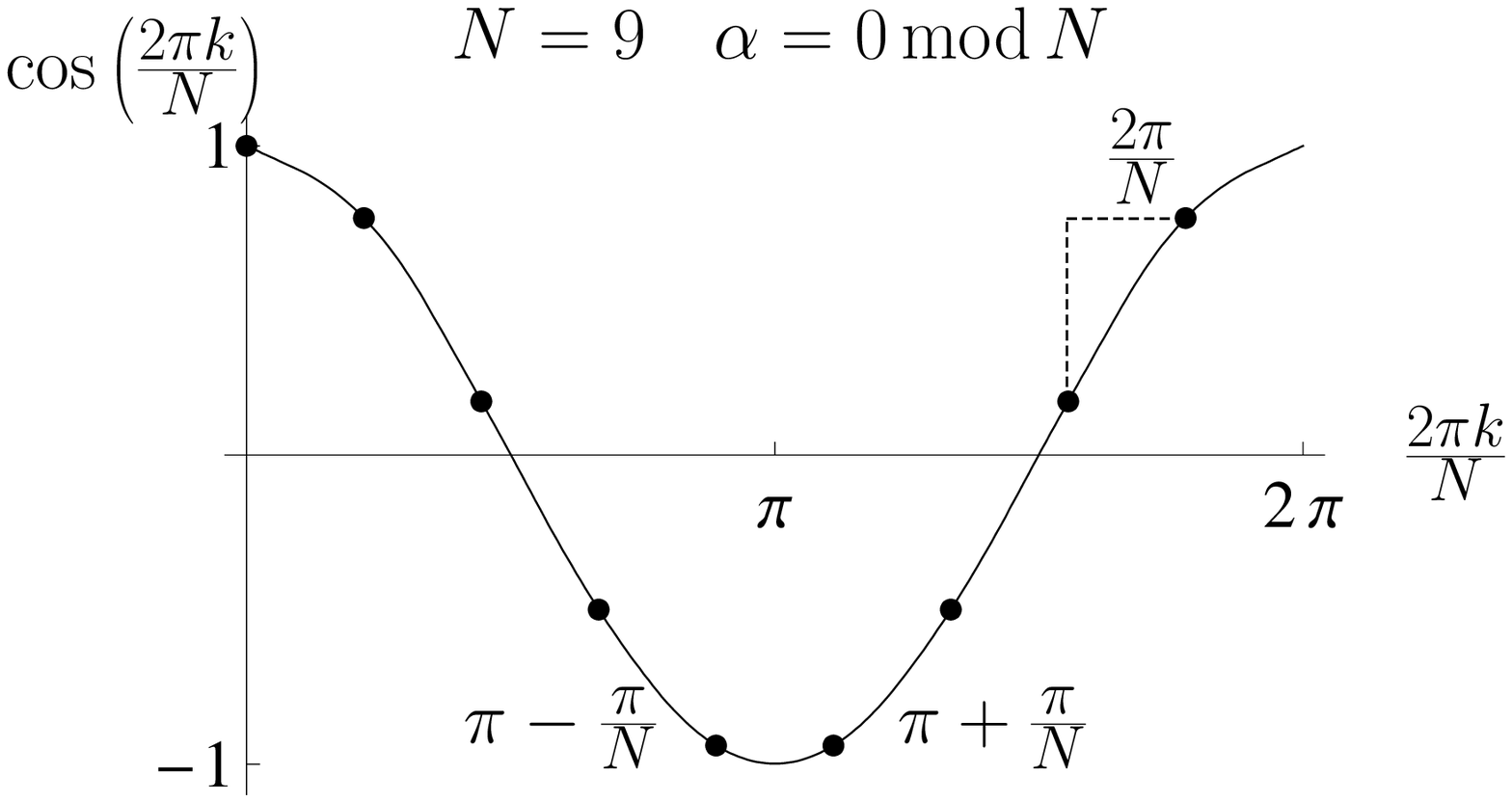}
&
\includegraphics[width=0.4\columnwidth]{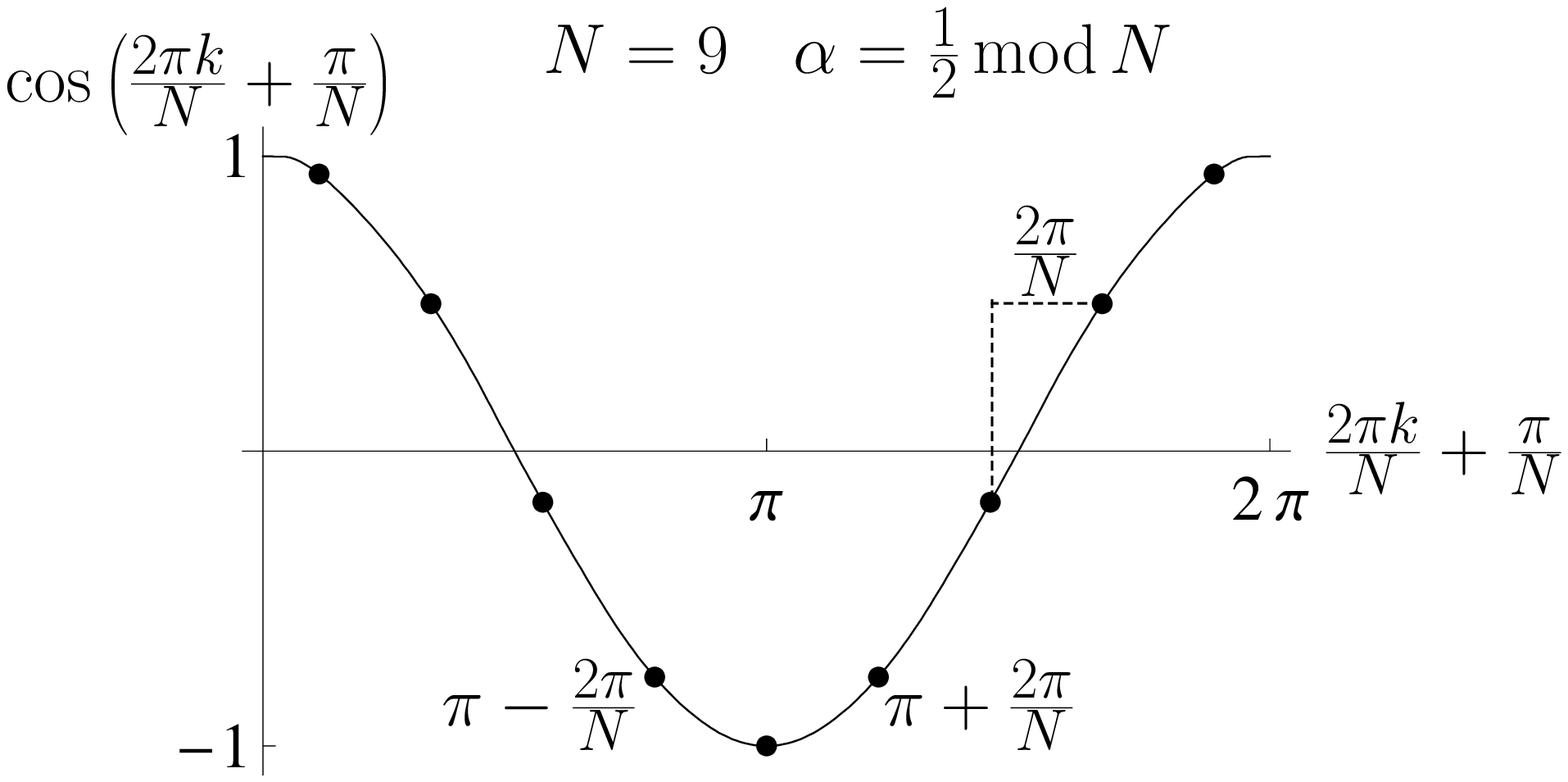}
\end{tabular}
\end{center}
\caption{Plot of
$\cos[2\pi(\alpha+k)/N]$, $k\in\mathbb{Z}_N$ for $N=8,9$ and
$\alpha=0\ \text{mod}\ N$, $\frac{1}{2}\ \text{mod}\ N$.}\label{fig:coseni}
\end{figure}

\begin{figure}
\begin{center}
\includegraphics[width=0.5\columnwidth]{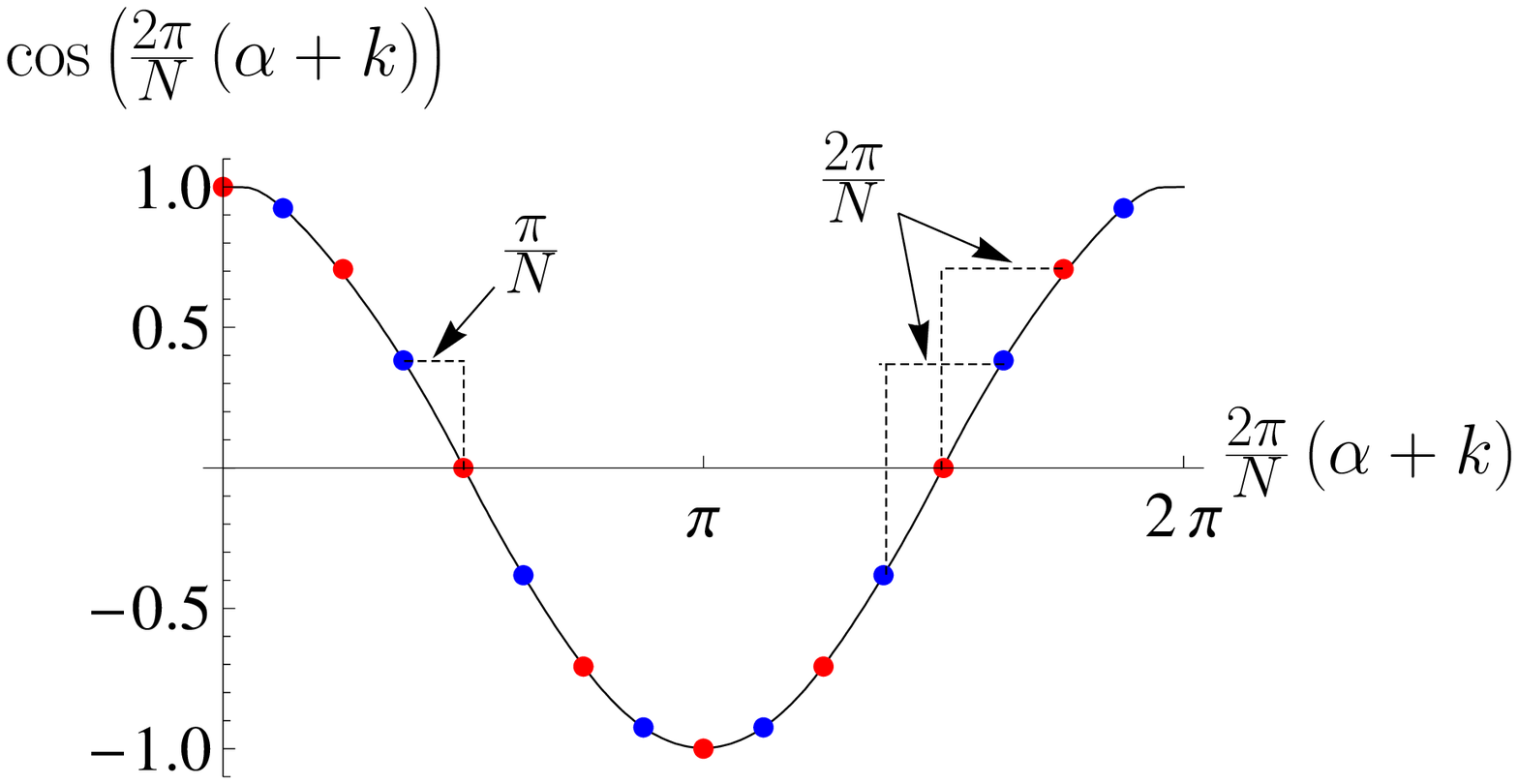}
\qquad
\includegraphics[width=0.4\columnwidth]{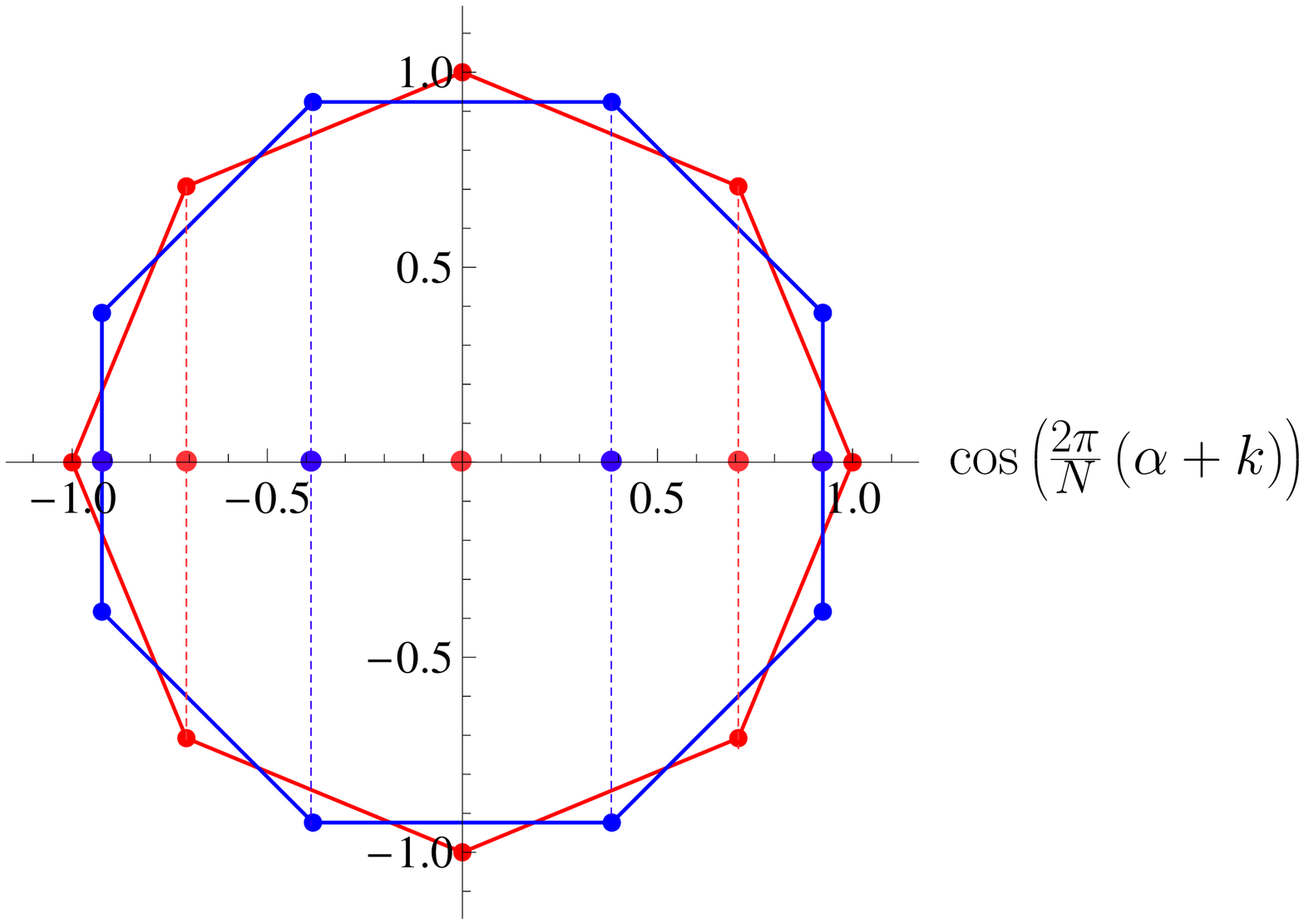}
\end{center}
\caption{(Color online) Left: plot of
$\cos[2\pi(\alpha+k)/N]$, $k\in\mathbb{Z}_N$ with $\alpha=0\ \text{mod}\ N$ (red) and
$\alpha=\frac{1}{2}\ \text{mod}\ N$ (blue), for $N=8$. Right:
geometrical description of
$\cos[2\pi(\alpha+k)/N]$.}
\label{fig:confronti a fissatoN}
\end{figure}

In this section we will focus on the spectrum of the system. We
shall set henceforth $J=1$. The state with no fermions has an
energy density that is independent of $\alpha$ and
of the number of sites $N$. Indeed, from Eq.\ (\ref{eq:newhamdiag}),
\begin{equation}\label{eq:enervacuum}
\varepsilon_\text{vac}(g)=\frac{E_\text{vac}(g)}{N}=\frac{1}{N}\sum_{k=0}^{N-1}\left[
g-\cos\!\left(2\pi
\frac{\alpha+k}{N}\right)
\right]=g.
\end{equation}
If we add one fermion the energy density reads
\begin{equation}\label{eq:singleenergy}
\varepsilon_1(k,g)=\varepsilon_\text{vac}(g)-\frac{2}{N}\left[
g-\cos\!\left(2\pi\frac{\alpha+k}{N}\right)
\right],
\end{equation}
where $\alpha$ depends on $N$:
\begin{equation}
\left\{
\begin{array}{ll@{\quad}l@{\quad}l@{}l}
N\;\mbox{even}&\Rightarrow N-1\;\mbox{odd}&\Rightarrow&
\alpha=0&\ \text{mod}\ N, \\
N\;\mbox{odd}&\Rightarrow N-1\;\mbox{even}&\Rightarrow&
\alpha=\frac{1}{2}&\ \text{mod}\ N.
\end{array}
\right.
\end{equation}
In Fig.\ \ref{fig:spettri singola particella} we represent the
single particle energy spectra corresponding to $N=8$ and $9$ sites
(representative of an even/odd number of fermionic sites,
respectively). Different curves are parametrized by
$k\in\mathbb{Z}_N$ and one notices the presence of degeneracies in
both cases. Since we are interested in the ground state of the
system, we focus on the lowest energy levels and consider the values
assumed by the function $\cos[2\pi(\alpha+k)/N]$ in the four
possible cases [2 parities of number of sites $N$ and 2 values of
$\alpha$ as in Eq.\ (\ref{alphadef})], as shown in Fig.\
\ref{fig:coseni}. Notice that these results can be described in
terms of regular polygons inscribed into a circle of unit radius,
see Fig.~\ref{fig:confronti a fissatoN}.

\begin{figure}
\begin{center}
\includegraphics[width=0.5\columnwidth]{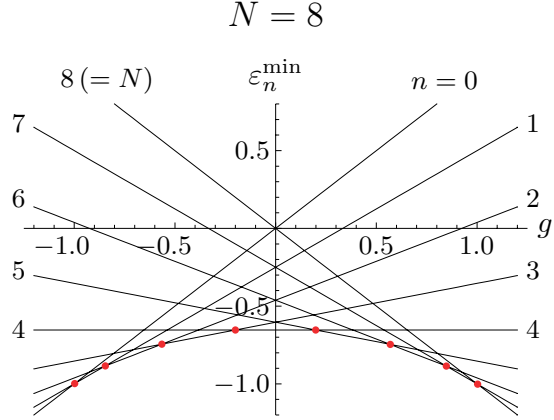}
\end{center}
\caption{Lowest-energy levels $\varepsilon_n^\text{min}(g)$ for different number
of fermions $n$; the intersection between the energy levels
corresponding to $n$ and $n+1$ fermions (starting from $n=0$) are
the quantum critical points (dots).}\label{fig:puntidintersezione}
\end{figure}

From  Fig.\ \ref{fig:coseni} one obtains
the values of $k$ that minimize the energy per site; in the 1-particle sector
one has
\begin{equation}\label{impulso 1 particella}
\left\{
\begin{array}{l@{\quad}l@{\quad}l@{}l}
\medskip
N\;\text{even}&\Rightarrow&
\alpha=0&\ \text{mod}\ N\Rightarrow k=\frac{N}{2},\\
N\;\text{odd}&\Rightarrow&
\alpha=\frac{1}{2}&\ \text{mod}\ N\Rightarrow
k=\frac{N-1}{2}.
\end{array}
\right.
\end{equation}
Similarly, in the 2-particle sector the minimum energy is at
\begin{equation}
\left\{
\begin{array}{l@{\quad}l@{\quad}l@{}l}
\medskip
N\;\mbox{even}&\Rightarrow&\alpha=\frac{1}{2}&\ \text{mod}\ N\Rightarrow
\{k_1,k_2\}=\left\{\frac{N}{2}-1,\frac{N}{2}\right\}, \\
N\;\mbox{odd}&\Rightarrow&\alpha=0&\ \text{mod}\ N\Rightarrow
\{k_1,k_2\}=\left\{\frac{N-1}{2},\frac{N+1}{2}\right\}.
\end{array}
\right.
\end{equation}
As a result, the general expression of the lowest energy levels in the different
$n$-particle sectors  does not depend on the
parity of $N$. For $n$ fermions, one gets
\begin{equation}\label{livelli minima energia}
\varepsilon_n^{\text{min}}(g)=g \left(1-\frac{2n}{N}\right) -
\frac{2}{N}\frac{\sin(n\pi/N)}{\sin(\pi/N)}.
\end{equation}
In Fig.\ \ref{fig:puntidintersezione} we plot the lowest energy
levels corresponding to $0\leq n \leq N$ for $N=8$ sites. The
intersections of levels corresponding to $n$ and $n+1$ fermions
(starting from $n=0$) define the \textit{level crossing points} or
\textit{quantum critical points} $g_c$, where an excited level
and the ground state are interchanged. The analytic expression of
the critical points is easily obtained by the condition
$\varepsilon_n^{\text{min}}(g_c)=\varepsilon_{n+1}^{\text{min}}(g_c)$.
We find
\begin{equation}\label{eq:intersectpoint}
g_c(n)=\frac{ \sin(n\pi/N)-\sin[(n+1)\pi/N]}{\sin(\pi/N)}, \qquad 0\leq n\leq N.
\end{equation}
As a consequence, the ground-state energy per site is
\begin{equation}\label{eq:enermin}
\varepsilon_\text{gs}(g)=g \left(1-\frac{2n}{N}\right) -
\frac{2}{N}\frac{\sin(n\pi/N)}{\sin(\pi/N)}\quad \text{with}\quad g\in ( g_c(n-1),g_c(n)),
\end{equation}
with $0\leq n\leq N+1$ and $g_c(-1)=-\infty$, $g_c(N+1)=+\infty$.
Thus, for $g\in(g_c(n-1),g_c(n))$, the ground state contains $n$ fermions. Note that $g_c(0)=-1$ and $g_c(N)=+1$, independently on $N$.

In the thermodynamic limit the critical points correspond to the
QPT\@. In order to study the properties of the ground state in this
limit, we evaluate the envelope of the lowest-energy levels. This is
obtained by the equation
\begin{equation}\label{eq: n per inviluppo}
\frac{\partial\varepsilon_n^\text{min}(g)}{\partial n}=0\quad\Rightarrow\quad
n(g)=\frac{N}{\pi}\arccos(-g\chi_N) ,
\end{equation}
where
\begin{equation}
\chi_N\equiv\frac{\sin(\pi/N)}{\pi/N}
\end{equation}
is a \textit{finite-size parameter} depending on the number $N$ of
sites in the chain.  The envelope is obtained by plugging $n(g)$
from Eq.\ (\ref{eq: n per inviluppo}) into Eq.\ (\ref{eq:enermin}).
By noting that  $n=0$ for $g\le-1/\chi_N$ while $n=N$ for
$g\ge1/\chi_N$, one gets
\begin{equation}\label{inviluppo vero del ground state al finito}
\varepsilon_\text{env}(g)=
\begin{cases}
\medskip
\displaystyle
g\left(1-\frac{2}{\pi}\arccos(-g\chi_N)\right)
-\frac{2}{\pi}\frac{\sqrt{1-g^2\chi_N^2}}{\chi_N}&
\displaystyle
\text{for}\quad |g|<\frac{1}{\chi_N} , \\
-|g|&
\displaystyle
\text{for}\quad
|g|>\frac{1}{\chi_N}.
\end{cases}
\end{equation}
It is easy to see that the envelope and its
first derivative are continuous functions at
$|g|=1/\chi_N$, whereas the second derivative
\begin{equation}
\frac{\text{d}^2\varepsilon_\text{env}(g)}{\text{d}g^2}=
\begin{cases}
\medskip
\displaystyle
-\frac{2}{\pi}\frac{\chi_N}{\sqrt{1-g^2\chi_N^2}}&
\displaystyle
\text{for}\quad
 |g|<\frac{1}{\chi_N},\\
 \displaystyle
0&
\displaystyle
\text{for}\quad|g|>\frac{1}{\chi_N}.
\end{cases}
\end{equation}
diverges to $-\infty$, for  $|g|\uparrow 1/\chi_N$.

\begin{figure}
\begin{center}
\includegraphics[width=0.4\columnwidth]{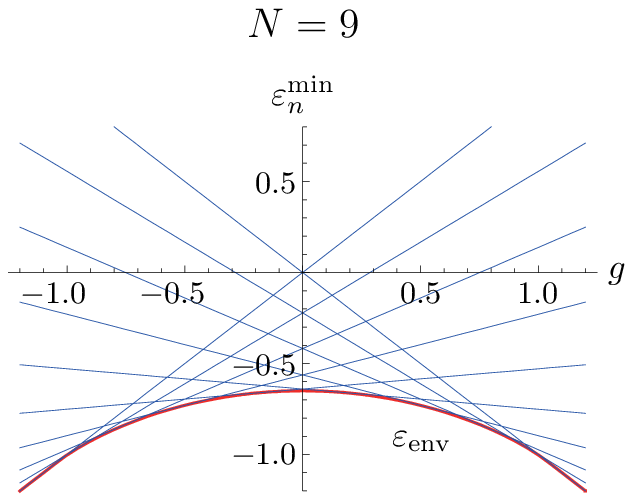}
\qquad
\includegraphics[width=0.4\columnwidth]{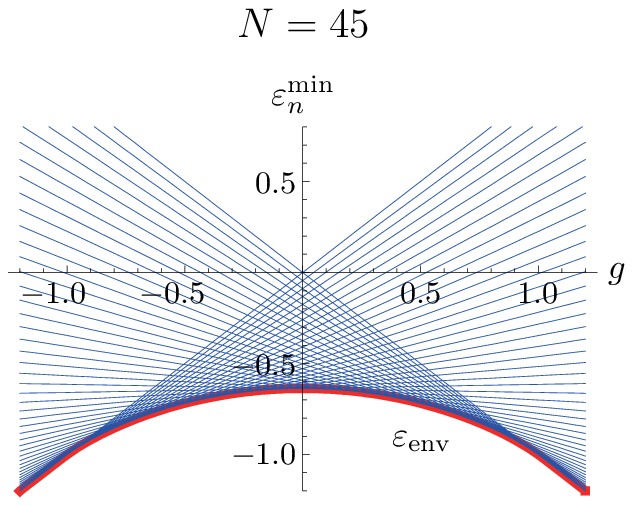}
\end{center}
\caption{Ground-state envelopes $\varepsilon_\text{env}(g)$ for $N=9$ and $45$
sites.}\label{fig:rette ed inviluppi per 9 e 45 siti}
\end{figure}

\begin{figure}
\begin{center}
\includegraphics[width=0.5\columnwidth]{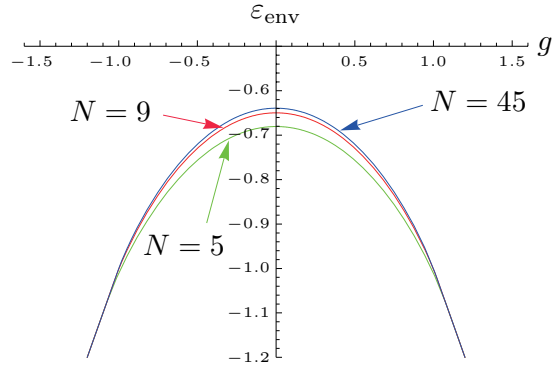}
\end{center}
\caption{Ground-state envelopes $\varepsilon_\text{env}(g)$ for $N=5$, $9$, and $45$
sites.}\label{fig:inviluppo5e9e45siti}
\end{figure}

In Fig.\ \ref{fig:rette ed inviluppi per 9 e 45 siti} we plot the
lowest-energy levels and the ground-state envelopes for $N=9$ and
$45$ sites.
Notice that the scales of the energy and magnetic field are similar.
This result is confirmed in Fig.\
\ref{fig:inviluppo5e9e45siti}, where we plot the envelopes for $N=9$
and $45$; their difference is negligible if compared to the envelope
for $N=5$, which shows that $N=9$ is already a good approximation of
the thermodynamical limit.

\begin{figure}
\begin{center}
\begin{tabular}{cc}
&
\includegraphics[height=0.29\columnwidth]{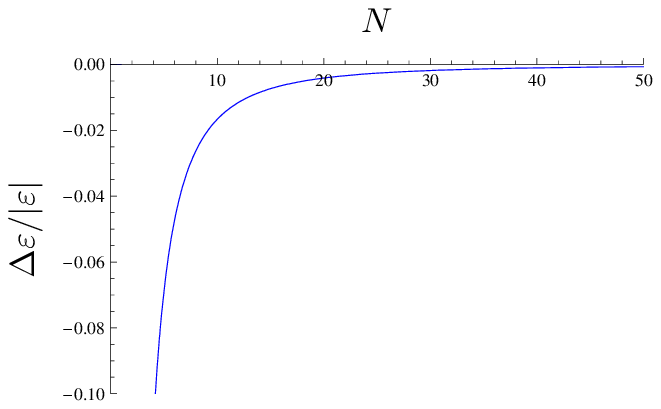}\\[-38.0mm]
\includegraphics[height=0.303\columnwidth]{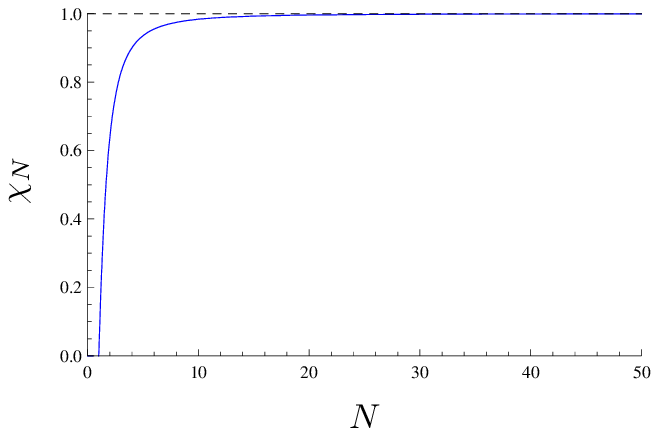}&
\end{tabular}
\end{center}
\caption{Finite-size parameter $\chi_N$ (left) and relative
error $\Delta\varepsilon/|\varepsilon|$ (right) for $N=1,\ldots,50$.}
\label{fig:raggiungimento limite termodinamico}
\end{figure}

It is interesting to evaluate the finite-size effects. This can be
done by considering the dependence of $\chi_N$ on $N$. We plot this
function in Fig.\
\ref{fig:raggiungimento limite termodinamico} (left). For
$N >10$, $\chi_N\simeq 1$ with excellent approximation. We
can obtain further information by looking at the thermodynamic limit
of the envelope (\ref{inviluppo vero del ground state al finito}),
\begin{equation}
\varepsilon_\infty(g)=\lim_{N\rightarrow\infty}\varepsilon_\text{env}(g)=
\begin{cases}
\medskip
\displaystyle
g\left(1-\frac{2}{\pi}\arccos(-g)\right) - \frac{2}{\pi}
\sqrt{1-g^2}&
\text{for}\quad |g|\leq1,\\
-|g|&
\text{for}\quad |g|\geq 1,
\end{cases}
\end{equation}
which is nothing but the ground state energy density of the thermodynamic system.
The relative error reads (for $g=0$)
\begin{equation}
\frac{\Delta\varepsilon}{|\varepsilon|}=\frac{\varepsilon_\text{env}(0)-\varepsilon_\infty(0)}{|\varepsilon_\infty(0)|}
=-\left(
\frac{1}{\chi_N}-1
\right)
\sim -\frac{1}{3!}\left(\frac{\pi}{N}\right)^2
\quad\text{for}\quad N\to\infty.
\label{eqn:error}
\end{equation}
It is interesting to notice that the relative error is $O(1/N^2)$,
which is better than what one naively expects, see comment after
Eq.\ (\ref{eq:aaa}).

Figure \ref{fig:raggiungimento limite termodinamico} (right)
confirms this result. The effect of the finite size of the chain can
be neglected already for a relatively small number of sites.

\subsection{Ground state revisited}
It is interesting to reinterpret the ground state of the chain in
terms of spins. For $g \leq -1$ the ground state is the state with
no fermions. By definition, this corresponds to all spins down
(antiparallel to the $z$ axis) in the chain
\begin{equation}\label{vuoto}
|\psi_0\rangle=|{\downarrow}\rangle|{\downarrow}\rangle
\cdots|{\downarrow}\rangle.
\end{equation}
This means that for strong magnetic fields the interaction between
the field and the spins is larger than that among the spins in the
$xy$ plane.

As the magnetic field tends to zero, the
number of spins in the down state decreases and the general expression of the
ground state for $g\in[g_c(n-1),g_c(n)]$ is given by
\begin{align}
|\psi_n\rangle=\frac{1}{N}\sum_{j_1<j_2<\cdots<j_n}
                       \Big[&\lambda_{j_1,j_2,\ldots,j_n}\sigma_{j_1}^+\sigma_{j_2}^+\cdots\sigma_{j_n}^+(\sigma_0^z)^n(\sigma_1^z)^n\cdots(\sigma_{j_1-1}^z)^n
\nonumber\\
&{}\times
(\sigma_{j_1}^z)^{n-1}(\sigma_{j_1+1}^z)^{n-1}\cdots(\sigma_{j_2-1}^z)^{n-1}
(\sigma_{j_2}^z)^{n-2}(\sigma_{j_2+1}^z)^{n-2}\cdots(\sigma_{j_3-1}^z)^{n-2}
\nonumber\\
&\hspace*{67mm}
{}\cdots\sigma_{j_n-1}^z\Big]|{\downarrow}\rangle_0|{\downarrow}\rangle_1\cdots|{\downarrow}\rangle_{N-1}\nonumber\\
=\frac{1}{N}\sum_{j_1<j_2<\cdots<j_n}
\Big[&\lambda_{j_1,j_2,\ldots,j_n}(-1)^{n j_1}(-1)^{(n-1)
(j_2-j_1)}(-1)^{(n-2) (j_3-j_2)}\cdots
(-1)^{j_{n}-j_{n-1}}\Big]\nonumber\\
&\hspace*{45mm}
{}\times
|{\downarrow}\rangle_0\cdots|{\uparrow}\rangle_{j_1}\cdots|{\uparrow}\rangle_{j_2}\cdots|{\uparrow}\rangle_{j_n}
\cdots|{\downarrow}\rangle_{N-1},
\end{align}
where $\lambda_{j_1,j_2,\ldots,j_n}$ is the sum over all permutations
$\{1,2,\dots,n\}\to\{p_1,p_2,\dots,p_n\}$
\begin{equation}
\lambda_{j_1,j_2,\ldots,j_n}=\sum_{p}(-1)^p
e^{\frac{2\pi\text{i}}{N}(k_1j_{p_1}+k_2j_{p_2}+\cdots
+k_nj_{p_n})}.
\end{equation}
Note that in this case the state of the system cannot be written
as a tensor product and $|\psi_n\rangle$ is entangled. When the magnetic
field is very strong and positive $g \geq 1$ (parallel to the $z$
axis) all  spins are aligned along its direction and the state of the chain is
\begin{equation}
|\psi_N\rangle=|{\uparrow}\rangle|{\uparrow}\rangle\cdots|{\uparrow}\rangle.
\end{equation}
These properties are summarized in Fig.\ \ref{fig: stati entanglati
e separabili}. We will analyze the entanglement structure for a
finite-size chain in the following section.
\begin{figure}[b]
\begin{center}
\includegraphics[width=0.5\columnwidth]{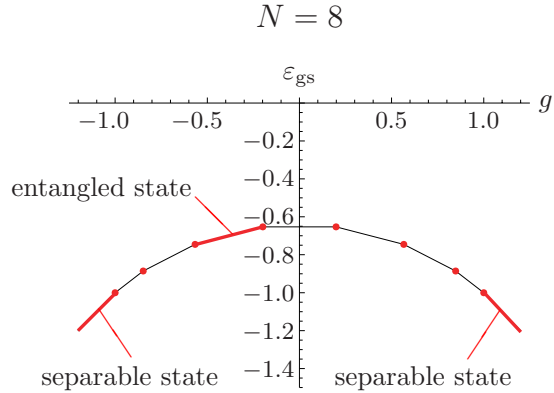}
\end{center}
 \caption{Classification of the ground states in terms of entanglement
 ($N=8$).}\label{fig: stati entanglati  e separabili}
\end{figure}

\section{Multipartite entanglement of the ground state}
\subsection{Probability density function characterization of
multipartite entanglement}
The evaluation of the entanglement stored in the ground states of
quantum spin chains has become a central problem in this field of
research. The study of the entanglement properties of the ground
state usually explores the link between quantum phase transitions
and entanglement generation and previous research mainly focused on
the thermodynamic limit \cite{QPT}.

One can use several measures of entanglement \cite{multipart}. A
typical approach consists in considering a given bipartition and
evaluating the bipartite entanglement (in terms of the von Neumann
entropy, the linear entropy, the purity, or in general a well
defined entanglement monotone \cite{sarorev}). A possible approach
to multipartite entanglement is to analyze the statistical
properties of bipartite entanglement over all balanced bipartitions
(each part consisting of one half of the chain) \cite{FFP}. This
characterization has been applied to the 1D quantum Ising model in a
transverse field, yielding interesting results
\cite{CFFP}.

We will characterize in a similar way the multipartite entanglement
of the XX chain. Consider a chain of $N$ spins and consider a
partition in two subsystems $A$ and $B$, made up of $N_A$ and $N_B$
qubits ($N_A+N_B=N$), respectively. For definiteness we assume $N_A
\le N_B$. The total Hilbert space is the tensor product
$\mathcal{H}=\mathcal{H}_A\otimes\mathcal{H}_B$ with dimensions
$\dim \mathcal{H}_A=2^{N_A}$, $\dim
\mathcal{H}_B=2^{N_B}$, and  $\dim \mathcal{H}=2^{N_A+N_B}=2^N$.

Let us denote the ground state by $|\psi_\text{gs}\rangle$ and
consider the purity of subsystem $A$ (which equals that of subsystem $B$)
\begin{equation}\label{eq:NABdefs}
\pi_{AB}(|\psi_\text{gs}\rangle)=\Tr_A \rho_A^2,\qquad
\rho_A=\Tr_B\rho,
\qquad
\rho=|\psi_\text{gs}\rangle\langle\psi_\text{gs}|,
\end{equation}
$\Tr_A$ ($\Tr_B$) being the partial trace over subsystem $A$ ($B$),
and take it as a measure of the bipartite entanglement between $A$ and
$B$\@.
We note that
\begin{equation}\label{eq:propNAB}
2^{-N_A}\leq
\pi_{AB}\leq 1 ,
\end{equation}
where the minimum (maximum) value is obtained for a completely mixed
(pure) state $\rho_A$. Therefore, a smaller value of $\pi_{AB}$
corresponds to a more entangled bipartition $(A,B)$. $\pi_{AB}$ will
depend on the bipartition and entanglement will be distributed in a
different way among all possible bipartitions. The key idea is that
the average of $\pi_{AB}$, denoted by $\mu$, measures the amount of
entanglement (the smaller $\mu$, the larger the entanglement),
whereas the standard deviation $\sigma$ measures how well this
entanglement is distributed among bipartitions. Clearly, if the
distribution function $\pi_{AB}$ is not very regular, higher moments
will be necessary in order to properly characterize it.

\subsection{Average and standard deviation of the distribution}
We numerically evaluated the distribution of bipartite entanglement
for the finite-size XX model on the circle. In Fig.\ \ref{fig:
mediatutti} we plot the average entanglement $\mu$ over balanced
bipartitions versus the coupling $g$  (for $N = 4,\ldots,10$ sites).
\begin{figure}
\begin{center}
\includegraphics[width=0.47\columnwidth]{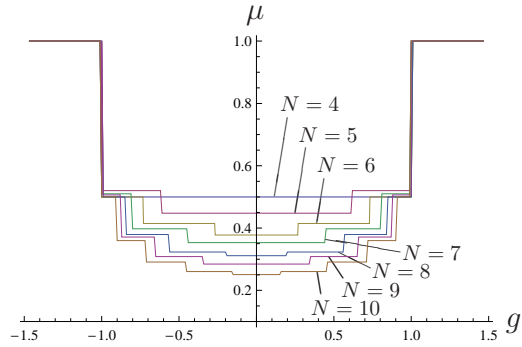}
\end{center}
\caption{(Color online) Average $\mu$ of $\pi_{AB}$ over all
balanced bipartitions for the XX chain with periodic boundary
conditions, with $N=4,\ldots,10$. }\label{fig: mediatutti}
\end{figure}
\begin{figure}
\begin{center}
\includegraphics[width=0.9\columnwidth]{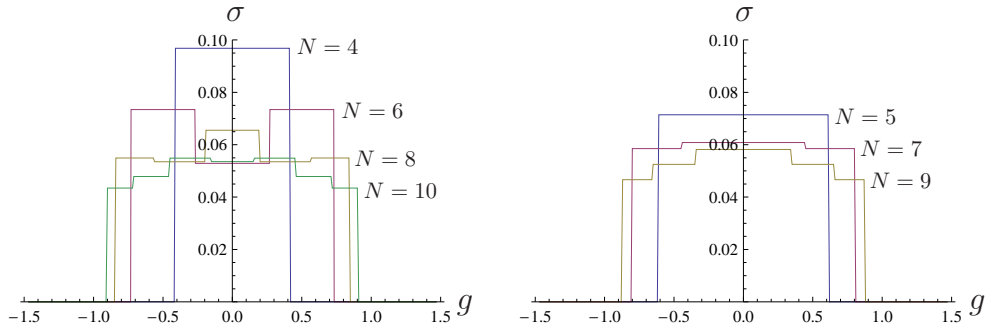}
\end{center}
\caption{(Color online) Standard deviation $\sigma$ of $\pi_{AB}$
over all balanced bipartitions for the XX chain with periodic
boundary conditions, with an even (left) and an odd (right) number
of sites. (Left) $N=4,6,8,10$. (Right) $N=5,7,9$.}\label{fig:
deviazioni}
\end{figure}

Note that, as expected, the maximum  entanglement is reached at
$g=0$. Moreover, the larger the system, the smaller $\mu$. An
interesting peculiarity of the XX model, at variance with the Ising
chain \cite{CFFP}, is that  $\mu(g)$ is not continuous: we observe
jumps between plateaux. These jumps correspond to the level crossing
points defined in Eq.\ (\ref{eq:intersectpoint}). In particular, for
$|g|>1$, the ground state of the model is completely factorized and,
consequently, we have no entanglement ($\mu=1$). The largest jump is
found at $|g|=1$, where $\mu \simeq 0.5$ for all $N$.

The standard deviation of the distribution is plotted in Fig.\
\ref{fig: deviazioni}. We displayed separately the cases $N$
even (left) and odd (right). Also in this case we observe a stepwise
behavior, depending on the number of fermions in the ground state.
For odd $N$ the maximum of $\sigma$ tends to decrease for a larger
chain: the curves are monotone for $-1\le g\le 0$ and $0\le g\le 1$.
On the other hand, for  even $N$, $\sigma$ shows a more complicated
structure; the curves are not monotone. The maximum is not a
decreasing function of the size (for $N=8$ it is larger than for
$N=6$). As a general trend, however, for both even and odd $N$,
$\sigma$ tends to decrease with $N$. Additional investigation is
required in order to understand the features of multipartite
entanglement and in particular the large-$N$ limit. As explained in
\cite{CFFP}, different scenarios are possible, depending on the
behavior of $\sigma/\mu$: if this ratio tends to zero with $N$,
entanglement will tend to be well distributed among different
bipartitions.

\section{Conclusions}
We analyzed the finite-size XX model on the circle. We diagonalized
the Hamiltonian by defining a deformed Fourier transform that takes
into account the periodic boundary conditions. We also studied the
ground state of the chain, by focusing on the points of level
crossings, that forerun the QPT in the thermodynamic limit, and
defined a finite-size parameter that quantifies how rapidly this
limit is approached. Finally, we looked at the properties of the
multipartite entanglement of the ground state in terms of the
distribution of bipartite entanglement. There is considerable
interest in the study of entanglement for quantum spin chains, both
in view of applications and because of their fundamental interest.
Future activity will focus on the study of different models, with
more general interactions and/or boundary conditions.

\vspace*{1cm}
{\bf Acknowledgements} This work is supported by the European
Community through the Integrated Project EuroSQIP and by the
bilateral Italian Japanese Projects II04C1AF4E on ``Quantum
Information, Computation and Communication" of the Italian Ministry
of Instruction, University and Research\@.

\end{document}